\documentstyle[aps,floats,prb,preprint]{revtex} 
\begin{document}  \preprint{June 27, 1994} \draft
\title{Quantum phase transition \\
  in the frustrated Heisenberg antiferromagnet }
\author{  A. V. Dotsenko and O. P. Sushkov\cite{Budker} }
\address{ School of Physics, The University of New South Wales,
             Sydney 2052, Australia}
\date{cond-mat/9406106; submitted to Physical Review B}
\maketitle
\begin{abstract}
Using the $J_1$-$J_2$ model,
 we present a description of quantum phase transition
 from N\'{e}el ordered to the spin-liquid state
 based on the modified spin wave theory.
The general expression for the gap in the spectrum
 in the spin-liquid phase is given.
\end{abstract}

\tighten
There has been recently considerable interest in magnetically
 disordered states in quantum spin models.
Much of this interest stems from the connection of this problem to
 high-$T_c$ superconductivity.
The ground state of undoped compound has long range antiferromagnetic
  order.
It is well described by the Heisenberg model and has been studied by
 numerous methods.\cite{Manousakis}
However introducing a small number of holes leads to destruction of
 long range order.
The resulting state is still not fully understood.


Destruction of long range order can be studied by introducing some
 frustration into the Heisenberg model.
We will focus on the simplest possible model of such kind
 which is the $J_1$-$J_2$ model defined by
\begin{equation} \label{H}
  H = J_1 \sum_{\rm NN} {\bf S}_{\bf r} \cdot {\bf S}_{{\bf r}'}
    + J_2 \sum_{\rm NNN} {\bf S}_{\bf r} \cdot {\bf S}_{{\bf r}'}.
\end{equation}
In this Hamiltonian, the $J_1$ term describes
 the usual Heisenberg interaction of nearest neighbour spins
  ($S={1\over2}$) on a square lattice,
 while the $J_2$ term introduces a frustrating interaction
  between next nearest neighbour sites.
The $J_1$-$J_2$ model itself is hardly applicable to any real
 materials\cite{jj-bad} (although it was originally
 proposed\cite{suggested} to describe high-$T_c$ superconductors).
However, it is valuable for demonstrating how long range order
 can be destroyed.
For convenience, we set $J_1 = 1$ and denote $\alpha \equiv J_2 / J_1$
 (the notation of Ref.\ \onlinecite{Manousakis} is used whenever possible).

For small $\alpha$, the ground state is N\'{e}el ordered.
For large $\alpha$ the system is decomposed into two N\'{e}el ordered
 sublattices which, however, have the same quantization axis.
This is the so-called collinear state.
Whether or not the N\'{e}el and collinear states are separated in
 parameter space by a state without long range order
 has been the subject of many discussions.
Besides the many spin wave
 calculations,\cite{linear,mean_field,spin-waves,Igarashi,Gochev}
 the model has been studied by
  the Schwinger boson mean field theory,\cite{Schwinger}
  analysis of small lattices,\cite{numerical}
  a series expansion,\cite{series}
  a mean field theory of bond operators,\cite{SB}
   and other methods.\cite{other}
Despite the numerous efforts, a strict answer has not been obtained.
Nevertheless, since only the mean field spin wave theory and
 the essentially equivalent  Schwinger boson mean field theory
 predict a first order transition from the N\'{e}el to the collinear state,
 while all other methods provide support for the existence of a
 different intermediate state, the latter scenario appears far
  more probable (this is also supported by a recent calculation
 of corrections to the mean field solution\cite{Gochev}).
In this work we assume that the system undergoes a second
 order quantum phase transition at a certain $\alpha = \alpha_c$ from
 the N\'{e}el to a spin-liquid state.

For the N\'{e}el phase we will use the spin wave theory
 which gives the following description.
The staggered magnetization
  $m^{\dag} = | \langle 0| S_{\bf r}^z |0 \rangle | $
 is equal to
\begin{equation} \label{sm}
m^{\dag} = {1\over2} - {2\over N} \sum_{\bf k} \sinh^2 \theta_{\bf k},
\end{equation}
where $N$ is the number of sites on the lattice, the summation is
performed over the Brillouin zone of one sublattice
 ($ |k_x| + |k_y| \le \pi$),
 and $\theta_{\bf k}$ is the parameter of the Bogolubov transformation
 determined in the linear spin-wave theory\cite{linear} (LSWT) by
\begin{equation} \label{theta}
 \tanh 2\theta_{\bf k} =
 {  \gamma_{\bf k}   \over  1 + \alpha (\eta_{\bf k}-1)    },
\end{equation}
or in the mean field spin wave theory theory\cite{mean_field} (MFSWT) by
 the selfconsistent solution of Eq.\ (\ref{sm}) together with
\begin{eqnarray} \eqnum{3$'$} \label{thetaMF}
&& \tanh 2\theta_{\bf k} =
 {
    (m^{\dag}+g_1) \gamma_{\bf k}
 \over
    (m^{\dag}+g_1) + \alpha (m^{\dag}+g_2) (\eta_{\bf k}-1)
 }, \\
&& g_1 = {2\over N} \sum_{\bf k} {1\over2}
              \gamma_{\bf k} \sinh 2\theta_{\bf k}, \ \
   g_2 = {2\over N} \sum_{\bf k}
               \eta_{\bf k} \sinh^2 \theta_{\bf k}. \nonumber
\end{eqnarray}
We have defined
\[
  \gamma_{\bf k} = {1\over2} (\cos k_x + \cos k_y)
 \ \  {\rm and} \ \
 \eta_{\bf k} = \cos k_x \cos k_y.
\]
Numerical values of $m^{\dag}$ in LSWT, MFSWT and
 other approximations are shown in Fig.\ \ref{SZ}.

Further, the dispersion of the Goldstone spin-wave excitations is
\begin{equation} \label{dispersion}
 \epsilon_{\bf k} =
\left\{
\begin{array}{ll}
2 \ (1 - \tanh^2 2\theta_{\bf k} )^{1/2}, & {\rm LSWT};\\
4 \ \biggl [ (m^{\dag}+g_1) + \alpha (m^{\dag}+g_2)
  (\eta_{\bf k} - 1) \biggr ]
  (1 - \tanh^2 2\theta_{\bf k} )^{1/2}, & {\rm MFSWT}.
\end{array} \right.
\end{equation}
for $k \ll 1$ is $\epsilon_{\bf k} = c k$.
We calculate $Z_c = c/c_0$
 ($c_0=\sqrt{2}$ is the spin wave velocity at $\alpha=0$ calculated by LSWT)
\begin{equation} \label{Zc}
 Z_c = \left\{ \begin{array}{ll}
(1-2\alpha)^{1/2}, & {\rm LSWT;}\\
2\sqrt{2} (m^{\dag}+g_1) \left( {1\over2} - \alpha
 {m^{\dag}+g_2 \over m^{\dag}+g_1} \right)^{1/2}, & {\rm MFSWT.}
\end{array}\right.
\end{equation}
The numerical value of $Z_c$ is given in Fig.\ \ref{velocity}.
Note the remarkable agreement of the mean field solution and
 the $1/S$ expansion when the latter is converging.

Let us now examine the accuracy of the spin wave theory as the staggered
 magnetization $m^{\dag}$ decreases.
The spin wave theory starts with transforming the spin operators
 ${\bf S}_{\bf r}$ to bosonic operators  $a^{\dag}_{\bf r}$  and
 $a_{\bf r}$ using the Dyson-Maleev or Holstein-Primakoff transformation.
In either case,  $S^z_{\bf r} = {1\over2} - \hat{n}_{\bf r}$ with
 $\hat{n}_{\bf r} = a^{\dag}_{\bf r} a_{\bf r}$
  (a spin up sublattice is considered).
The physical states are those with $n_{\bf r} = 0$ and $1$.
However, the Hamiltonian is in fact simplified so that it connects
 the physical states to unphysical
 (see discussion in Ref.\ \onlinecite{Manousakis}).
To estimate the amount of unphysical states  introduced to the wave function
 we calculated averages of higher powers of the $\hat{n}_{\bf r}$ operator
\begin{equation} \label{powers}
\langle 0| \hat{n}_{\bf r} |0 \rangle  = m \equiv {1\over2} - m^{\dag},
 \ \
\langle 0| \hat{n}_{\bf r}^2 |0 \rangle  = m+2m^2, \ \
\langle 0| \hat{n}_{\bf r}^3 |0 \rangle  = m+6m^2+6m^3.
\end{equation}
 where $m$ is defined by Eq.\ (\ref{sm}).
If only $n_{\bf r} = 0$ and $1$ were present, we would have
  $\hat{n}^l_{\bf r} = \hat{n}_{\bf r}$ ($l \geq 1$).
The problem of unphysical states becomes serious as $m$ increases.
To estimate the weight of the states with $n_{\bf r} >1$,
 we proceed as follows.
Expand the state obtained in the spin wave theory in
 states with a definite number of bosonic excitations at site ${\bf r}$
\begin{equation} \label{expansion}
|0 \rangle = \sum_{n=0}^{\infty} c_n |n_{\bf r} \rangle,
\end{equation}
where $ \hat{n}_{\bf r} |n_{\bf r}\rangle = n_{\bf r} |n_{\bf r}\rangle.$
Obviously, the correlators of Eqs.\ (\ref{powers}) can now be expressed as
\begin{equation} \label{tr}
 \langle 0| \hat{n}_{\bf r}^l |0\rangle =
 \sum_{n=0}^{\infty} n^l c_n^2, \ \ l = 0,\ 1,\ 2,\ 3,\ldots
\end{equation}
Now we  truncate the series (\ref{expansion}) at $c_4$
 and with the truncated series we solve the first four of Eqs.\
 (\ref{tr}) taking the left-hand-side from Eq.\ (\ref{powers}).
The results are presented in Table \ref{table3}.
\begin{table}
\caption{ \label{table3}
The weight of unphysical states when the series (\protect\ref{expansion})
 is truncated at $c_4$.
}
\begin{tabular}{ldddd}
$m$ & $c_0^2$ & $c_1^2$ & $c_2^2$ & $c_3^2$ \\
\tableline
0.100&0.852&0.209& $-$0.074 &0.013\\
0.196\tablenote{The $m$ for $\alpha=0$}
  &0.738&0.350& $-$0.112&0.023\\
0.300&0.659&0.412&$-$0.100&0.029\\
0.400&0.635&0.357&$-$0.020&0.028\\
0.500&0.675&0.167& 0.142&0.017\\
\end{tabular}
\end{table}
Because a truncated series is used, negative weights sometimes appear.
However, for the purpose of estimating, the method is adequate.
We see that at $m=0.5$ unphysical states constitute
 about 20 \% of the wave function and we conclude that the spin wave
 approximation should  be quite reasonable even at this point.
(However, we must restrict ourselves to considering only low powers
 of the operators $a^{\dag}_{\bf r}$ and $a_{\bf r}$ since high powers
 have large contributions from unphysical states and may
 lead to absolutely unphysical results.)

We assume that the system undergoes a second order transition
 into a liquid state at $\alpha=\alpha_c$ (the point when the
 sublattice magnetization becomes zero).
In terms of the initial ($\alpha = 0$) ground state,
 zero sublattice magnetization means that
 the ground state is a condenstate of many spin waves
 $a_{\bf k}$ and $b_{\bf k}$.
To describe the emerging phase we must take into account their
 non-linear interaction.
We cannot do this exactly.
However, there is an approximate method using the suggestion
 made by Takahashi\cite{Takahashi}
 for the Heisenberg model at non-zero temperature.

Following Takahashi, we impose an additional condition that
 sublattice magnetization is zero
\begin{equation} \label{constraint}
\langle 0| S^z_A - S^z_B |0 \rangle =
 \langle 0|
 \case{1}{2} - a^{\dag}_{\bf r} a_{\bf r}  +
 \case{1}{2} - b^{\dag}_{\bf r} b_{\bf r} |0 \rangle
 = 0,
\end{equation}
where $A$ and $B$ are the spin up and down sublattices.
In essence it means an effective cut-off of unphysical states.
In fact, Eq.\ (\ref{constraint}) together with conservation
  of the $z$-component of the total spin is equivalent to
\begin{equation}
   \langle 0| \sum_{\bf k}  a^{\dag}_{\bf k} a_{\bf k} |0 \rangle
 = \langle 0| \sum_{\bf k}  b^{\dag}_{\bf k} b_{\bf k} |0 \rangle
 = {N \over 4}.
\end{equation}
The total number of single spin wave states in the magnetic
 Brillouin zone is $N/2$.
Therefore, after introducing Eq.\ (\ref{constraint}) the effective
 number of allowed states in the Hilbert space of the system is
\[
 \left(
{
 (N/2)!
\over
 (N/4)! (N/4)!
}
 \right) ^2
\sim {4 \over \pi N} 2^N,
\]
so that with logarithmic accuracy the correct dimensionality ($2^N$) is
 restored.

The constraint (\ref{constraint}) is introduced into the Hamiltonian
 via a Lagrange multiplier $\lambda$.
Now we we must diagonalize
\begin{equation} \label{new_H}
 H_{\lambda} = H - \lambda \left(
 \sum_{A} a^{\dag}_{\bf r} a_{\bf r} +
 \sum_{B} b^{\dag}_{\bf r} b_{\bf r}  \right).
\end{equation}
The simple (linear) second term here [taken together with
 Eq.\ (\ref{constraint})] takes account of non-linear interaction
 of spin waves.
Diagonalizing Eq.\ (\ref{new_H}) we get the spectrum of excitations which
 has a gap $\Delta \propto \sqrt{\lambda}$.
When $\Delta \ll 1$, the spectrum is
\begin{equation}
  E_{\bf k} = ( \Delta^2 + \epsilon^2_{\bf k} )^{1/2},
\end{equation}
 where $\epsilon_{\bf k}$ is the dispersion from Eq.\ (\ref{dispersion})
 and $\Delta$ is determined from
(for clarity, we present now equations only in the form they have in LSWT)
\begin{equation} \label{new_constr}
 m^{\dag} + 2 \int \! \! \int { d^2 {\bf k} \over (2\pi)^2 }
 \left( {1\over \epsilon_{\bf k} } - {1\over E_{\bf k}} \right) = 0
\end{equation}
[here $m^{\dag} <0 $ is calculated from Eq.\ (\ref{sm})].
The integral in Eq.\ (\ref{new_constr})
 converges at small $k$ where $\epsilon_{\bf k} \approx c k$.
After integration we have for $\alpha > \alpha_c$
\begin{equation}
 \Delta = \pi |m^{\dag} | c^2.
\end{equation}
The correlation length in this state is $\xi \propto 1/\Delta$.

In the vicinity of the transition
$\Delta = \pi c_c^2 A (\alpha - \alpha_c)$,
 where $c_c$ is the spin wave velocity at the critical point
 ($c_c \approx 0.71$ whether LSWT or MFSWT is used, see Fig.\ 2)
and $A = -\partial m^{\dag} /\partial \alpha$
 is the slope of dependance of $m^{\dag}$ versus $\alpha$ (Fig.\ 1).
We can estimate $A$ from LSWT or the improved MFSWT\cite{Gochev}
 and we obtain $\Delta \approx 3.3 (\alpha - \alpha_c)$.

It is easy to generalize the consideration to non-zero temperatures.
In this case the gap $\Delta$ will be determined from
\begin{equation}
m^{\dag} +   2 \int \! \! \int { d^2 {\bf k} \over (2\pi)^2 }
 \left[ {1\over \epsilon_{\bf k}} -
 {1\over E_{\bf k}} \coth \left( {E_{\bf k} \over 2T} \right) \right] = 0.
\end{equation}
Similarly to the $T=0$ case, the integration can be
 performed exactly for $T, \Delta \ll 1$ yielding
\begin{equation} \label{gap-eq}
{T \over 2\pi} \log \left( 2 \sinh {\Delta \over 2T} \right) = -\rho_s,
\end{equation}
where
\[ \rho_s =  {1\over4} m^{\dag} c^2. \]

Equation (\ref{gap-eq}) has been obtained in the $N_d = \infty$ limit
 of the non-linear $\sigma$-model\cite{Sachdev}
   ($N_d=3$ is the number of components of the order parameter)
 which describes long range behaviour of spin systems using several
 phenomenological parameters.
Solutions of Eq.\ (\ref{gap-eq}) in different regimes
 have been discussed\cite{Sachdev} (see also Ref.\
 \onlinecite{trieste}) and we will not dwell on this issue.

To summarize, we have presented a description of quantum melting
 of long range antiferromagnetic order in the frustrated Heisenberg
 model.
The suggested approach can be used for many systems,
 in particular to the doped $t$-$J$ model.\cite{Kuch}

\vskip10pt
We would like to thank F. Mila for drawing attention to the topic.
We are also grateful to C. Hamer, M. Yu.\ Kuchiev, D. Poilblanc,
 and T. Ziman for valuable discussions.
This work forms part of a project supported by a grant of the Australian
 Research Council.

\tighten

\newpage
\begin{figure} \caption{  \label{SZ}
Staggered magnetization $m^{\dag}$ as a function of $\alpha$
 in the N\'{e}el and collinear states.
Dotted line is the LSWT result,\protect\cite{linear}
solid line the MFSWT result,\protect\cite{mean_field}
long dashed line the improved MFSWT result,\protect\cite{Gochev}
dashed line obtained by the $1/S$ expansion\protect\cite{Igarashi}
 in the first ($\circ$) and second ($\bullet$) order.
}  \end{figure}

\begin{figure} \caption{  \label{velocity}
Spin wave velocity renormalization $Z_c$ as a function of $\alpha$.
Dotted line is the LSWT result,\protect\cite{linear}
solid line the MFSWT result,\protect\cite{mean_field}
dashed line obtained by the $1/S$ expansion\protect\cite{Igarashi}
 in the first ($\circ$) and second ($\bullet$) order.
}  \end{figure}

\end{document}